%% file: PointEval.tex
\documentclass{article}
\usepackage{graphicx}
\usepackage{booktabs}   %% For formal tables:
\usepackage{subcaption} %% For complex figures with subfigures/subcaptions
\usepackage{todonotes}
\usepackage{soul,color}
\usepackage{listings}
\usepackage{enumitem}

\usepackage{color}
\usepackage{multirow}
\usepackage{hyperref}
\usepackage{wrapfig,lipsum,booktabs}
\usepackage{cite}
\usepackage{amsmath,amssymb,amsfonts}
\usepackage{algorithmic}
\usepackage{graphicx}
\usepackage{textcomp}
\usepackage{xcolor}
\usepackage{adjustbox}
\usepackage[affil-it]{authblk}

\definecolor{pblue}{rgb}{0.13,0.13,1}
\definecolor{pgreen}{rgb}{0,0.5,0}
\definecolor{pred}{rgb}{0.9,0,0}
\definecolor{pgrey}{rgb}{0.46,0.45,0.48}

\usepackage{listings}
\newcommand{\etal}{et.~al.}

\begin{document}
\title{PointEval: On the Impact of Pointer Analysis Frameworks}
\author{Jyoti Prakash, Abhishek Tiwari, Christian Hammer\\ \texttt{\{jyoti,tiwari,chrhammer\}@uni-potsdam.de}}
\affil{University of Potsdam, Potsdam, Germany}
\maketitle              % typeset the header of the contribution
\begin{abstract}
\input{sections/abstract.tex}

% \keywords{Software engineering \and Java \and Program Analysis \and Pointer Analysis }
\end{abstract}
\section{Introduction}
\input{sections/introduction.tex}
\section{Background}
\input{sections/background.tex}

\section{Evaluation Setup}
\input{sections/methodology.tex}

\section{Evaluation on DaCapo}
\input{sections/evaluation.tex}
\section{Evaluation on PointBench}
\input{sections/evaluation-pb.tex}
%\section{Threats to Validity}
%\input{sections/discussion.tex}
\section{Related Work}
\input{sections/related_work.tex}
\section{Conclusion}
\input{sections/conclusion.tex}

\bibliography{PointEval.bib}

\end{document}

%% file: sections/abstract.tex
%!TEX root = ../PointEval.tex
Pointer analysis is a foundational analysis leveraged by various static analyses. Therefore, it
gathered wide attention in research for decades. Some pointer
analysis frameworks are based on succinct declarative specifications. However, these tools are
heterogenous in terms of the underlying intermediate representation (IR), heap abstraction, and
programming methodology. This situation complicates a fair comparison of these frameworks and
thus hinders further research. Consequently, the literature lacks evaluation of the strengths
and weaknesses of these tools. 

In this work, we evaluate two major frameworks for pointer analysis, WALA and Doop, on the DaCapo
set of benchmarks. We compare the pointer analyses available in Wala and Doop, and
conclude that---even though based on a declarative specification---Doop provides a better pointer analysis than Wala in terms of precision and
scalability. We also compare the two IRs used in Doop, i.e., Jimple from the Soot framework and IR
from the Wala framework. Our evaluation shows that in majority of the benchmarks Soot's
IR gives a more precise and scalable pointer analysis. Finally, we propose a micro-benchmark \emph{PointerBench}, for which we manually
validate the points-to statistics to evaluate the results of these tools. 
%\todo{results?}

%% file: sections/introduction.tex
%!TEX root = ../PointEval.tex
%\iffalse
%    Paragraph 1: Introduce the problem
%    Para 2: Why the problem is important
%    Para 3: Why do we need a benchmark study
%    Para 4: Contributions
%        4.1. Extensive study and comparison of three benchmarks
%        4.2. PointBench, a collection of jar files which handles corner cases in Pointer analysis
%        4.3. Study of same benchmark
%\fi

Pointer analysis is a technique to statically infer the objects referred by a variable in all possible executions. Being a fundamental static analysis problem, it has gathered wide attention in recent literature~\cite{Yannis2017-OOPSLA-PTaint, YannisOOPSLA2009,tamiflex2011Bodden}. However, pointer analysis is a long standing problem in static analysis. Some of the challenges discussed by Hind et al.~\cite{hind-2001-havent-solved-yet}, such as precise analysis within scalable time constraints, have been addressed by now~\cite{eichberg2018soap,Rief2019}. However, other problems, such as whole program analysis and dynamic properties of languages, have not been completely solved. Therefore, pointer analysis has gathered widespread attention in the program analysis community and researchers have leveraged various static analysis techniques such as CFL reachability~\cite{Sridharan-cfl} and IFDS~\cite{boomerang}.

Many static analysis frameworks offer built-in support for pointer analysis such as Soot~\cite{soot2019}, Wala~\cite{wala2019}, and Doop~\cite{doop2019}. Researchers use these existing approaches as a foundation for their analyses~\cite{hybdridroid,modular-typestate-Aldrich}. However, these frameworks are heterogenous in terms of number of features used to abstract programs such as intermediate representation (IR) of code, methods of modeling allocation sites, and representing heap objects. This complicates a fair comparison for comparing existing pointer analysis as it may impact precision. It also impacts the researcher who want to use pointer analysis. They often ask: (1) Which pointer analysis framework to use? (2) What impact will it have on the precision of an upstream analysis? and (3) How easy is it to integrate \emph{them} with an upstream analysis? Answer to these questions helps the user of pointer analysis to make an informed decision on the use of a framework. Although, there has been numerous efforts on improving pointer analysis, little to no work has been done to compare the available frameworks. In this paper, we bridge this gap by comparing the state-of-the-art pointer analysis frameworks and provide relevant insights into strengths and weakness of each framework. 

% It complicates a fair comparison of two different frameworks. Further, a user of pointer analysis always face a question (1) Which pointer analysis to use? (2) What shall be the level of scalability and (3) How easy it is integrate a pointer analysis? Answer to these questions can enable a static analysis developer/user to make an informed decision for the usuability of the analysis. 

\iffalse
    Motivation behind this paper
    1. Which pointer analysis to use?
    2. What shall be the level of scalability?
    3. How easy it is to integrate a pointer analysis

    Points to Write:
    ==============================================================
    1. 
\fi
%  It gives a comparison, colloquially called as
% comparison between "apples and oranges". Further, \todo{which}many of these frameworks lack
% standardized benchmark results, making it unusable for further research. 

To this end, we compare the two existing state-of-the-art pointer analysis frameworks: (1) Doop~\cite{yannis-doop-souffle-2017,YannisOOPSLA2009}, and (2) Wala~\cite{wala2019}. \textit{Doop} is based on a declarative specification while \textit{Wala} is an imperative static analysis framework. Both frameworks are used by researchers to implement their analysis. To study their impact, we evaluate Doop with the different front-ends and compare their results.
To understand the differences between the pointer analysis frameworks we also develop a microbenchmark \emph{PointBench}. Our evaluation shows that reproducing previously published results is as challenging as comparing different pointer analyses, even if they are based on the same frontend. In general, we observed that Doop is faster and more precise than Wala. Our contributions in this paper are:
% We also compare the results of Wala v/s Doop in terms of precision and scalability\todo{Write some real numbers as well}. 
\begin{itemize}
\item We perform an independent evaluation of the Doop framework on a set of independent benchmarks and discuss the differences with the previously published results. Our evaluation shows that the existing results are not reproducible with the average decrease in points-to size by (at least) half.
\item We compare the evaluation of Doop with Wala. Our results show that Doop is scalable than Wala for a precise analysis, such as a two-callsite. The two-callsite analysis on Doop terminates within six hours for nine out of 11 benchmarks, while Wala fails for each baechmark within 7 hours.
\item We compare the results of Doop with different IRs and study the differences. We observe that the choice of IR does not significantly affects the precision and scalability.
%We also compare the results of Doop and Wala using the same IR.
% \item We also compare the results of
% Soot and Wala analyses and comment on their \todo{???}equivalence in other frameworks. 
\item We also propose a micro-benchmark, \emph{PointerBench}, containing corner cases for pointer analysis, with which we provide an evaluation of these frameworks. Our evaluation shows that on some microbenchmarks both pointer analysis frameworks could not terminate even within 90 minutes. Both frameworks achieve a precision less than $20\%$ on these microbenchmarks. 
\end{itemize}

%% file: sections/background.tex
%!TEX root = ../PointEval.tex
%\todo{Write an intrductory sentence}
Pointer analysis is a technique to statically infer the objects referred to by a variable
in all possible executions of a program. In case a language (e.g. a \emph{While} language) only supports assignments and variable
definitions, then pointer analysis is simply computing the transitive closure over assignments.
However, for practical programming languages this is not the case, as we have operations such as
function invocations, field assignments, and more. Exactly modeling these features is undecidable~\cite{landi-undecidable,ramalingam-alias,Reps-cs-undecidable} and therefore, we
need approximations for a decidable pointer analysis.

\begin{lstlisting}[language=Java,caption={Heap Allocation program},label={program1:loop},float=tb,belowskip=-0.8\baselineskip]
(*\label{program1:1}*)x = new Object();//o1
while (condition) {
(*\label{program1:3}*)  x.f = new Object();//o2
  x = x.f;
}
\end{lstlisting}

\subsection{Intermediate Representation}
Generally, program analysis  do not directly use the actual source code but use some Intermediate Representation(IR). Therefore, various framework support some form of IR usually based on Three-address code or Single Static Assignment(SSA). Therefore, it raises a question on the effects of IR for program analysis.

\begin{figure}[tb]
  \centering
  \begin{subfigure}{0.45\linewidth}
    \centering
    \includegraphics[scale=0.5]{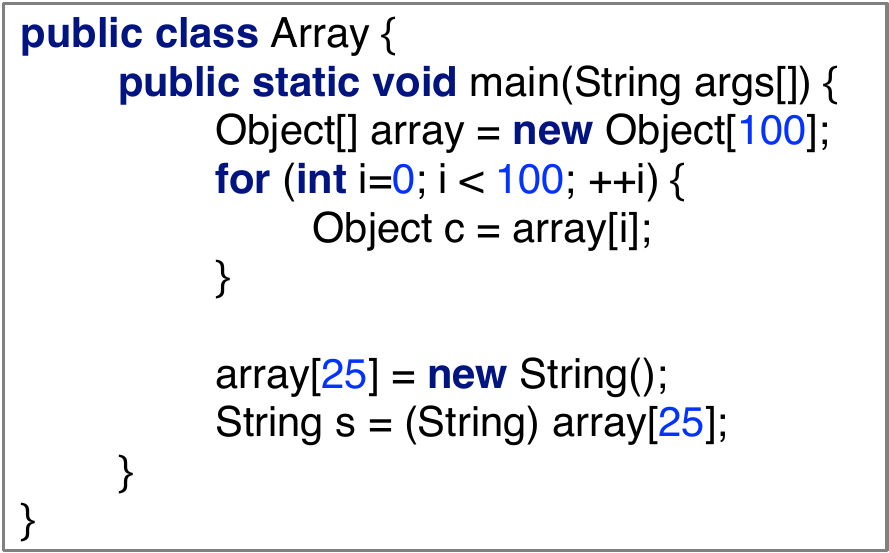}
    \caption{Java code from one of the microbenchmark}
    \label{fig:java-code}
  \end{subfigure}
  \begin{subfigure}{0.45\linewidth}
    \centering
    \includegraphics[scale=0.5]{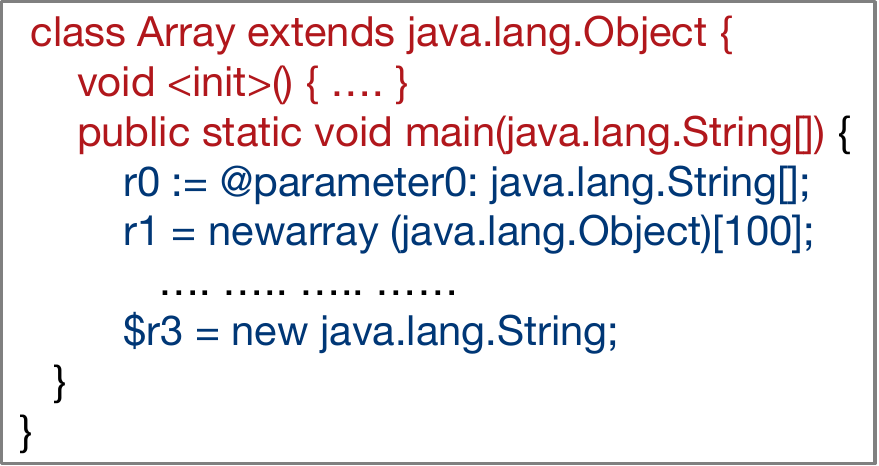}
    \caption{Jimple IR for code snippet in Fig~\ref{fig:java-code}}
    \label{fig:jimple-ir}
  \end{subfigure}
  \begin{subfigure}{0.9\linewidth}
    \centering
    \includegraphics[scale=0.5]{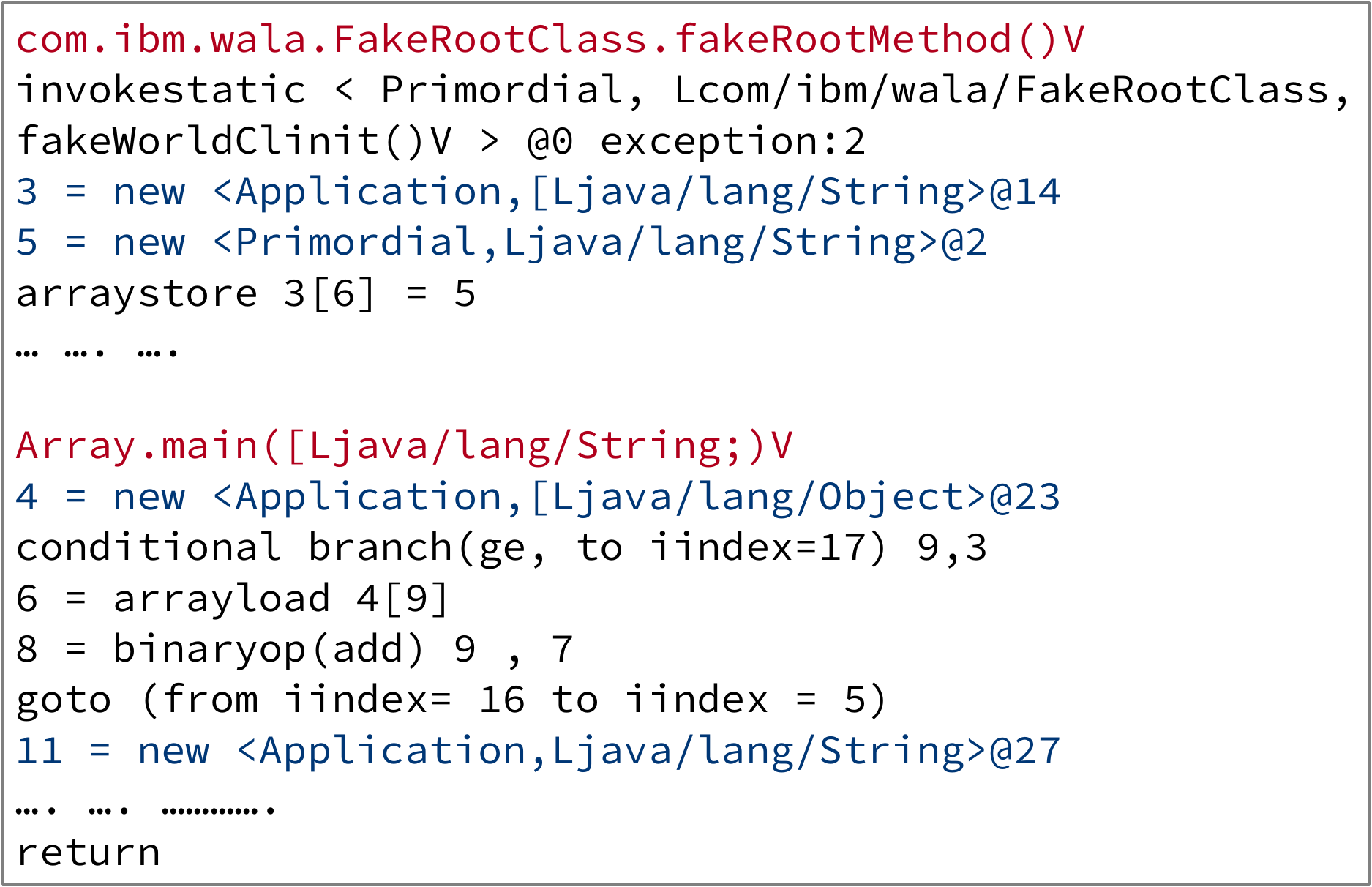}
    \caption{Wala IR for code snippet in Fig~\ref{fig:java-code}}
    \label{fig:wala-ir}
  \end{subfigure}
  \caption{Intermediate Representation }
  \label{fig:ir}
  \vspace{-2em}
\end{figure}

Wala supports Wala IR based on SSA and expressed as register transfer language(RTL). Soot supports multiple IRs but usually defaults to Jimple. Jimple is based on three address code and is also expressed as RTL. Being based on SSA Wala supports Pi Nodes which again creates a new variable while Jimple does not. However, apart from these syntactical differences there are differences int the use of IR for pointer analysis. In Wala, existing pointer analysis algorithms creates a set of synthetic methods (such as fakeRootMethod in Fig.~\ref{fig:wala-ir}) which is used a entry points and is connected to multiple entry points in a program. Each synthetic method creates some objects for initialization (cf. Fig.~\ref{fig:wala-ir}). Therefore, it reduces the precision score. However, this is not the case in Jimple where is does not create any synthetic objects~(cf.~Fig~\ref{fig:jimple-ir}). 
\subsection{Heap Abstraction}

One important aspect of pointer analysis is the heap abstraction~\cite{kanvar2016heap-abstraction}. A heap abstraction represents an object allocation symbolically. To bound the number of representations for allocations, a standard technique is \textit{allocation site abstraction},
which assigns a separate symbol for each allocation site in a program. For example, in Listing \ref{program1:loop}, the allocations on Line~\ref{program1:1} and Line~\ref{program1:3} are represented by different symbols \emph{o1} and \emph{o3}, the latter of which modeling all objects created within the loop body.     

\subsection{Pointer Analysis Frameworks}

\textbf{Doop} Doop~\cite{YannisOOPSLA2009} is a whole program pointer analysis framework. Doop is implemented
in Datalog, a logic programming language which supports declarative
specifications for many program analyses. The declarative specification consists of a set of ground facts
and logic rules. Logic rules are a set of predicates specified in a subset of first-order logic that
disallows complex terms such as functions in predicates~\cite{datalogWiki2019} (like
$pred(f(x),y)$). A Datalog specification reduces the overhead of writing boilerplate code, such that
static analysis designers can focus on specification and logic. \\% than writing boilerplate code such as data structures.
\textbf{Soot} Soot~\cite{soot2019} is an open-source general program analysis framework. Like
WALA, it supports fixed-point iteration and IFDS. It also supports two pointer analysis algorithms.
Unlike Wala, Soot supports four types of IR: \emph{Baf} is a streamlined representation of bytecode,
\emph{Jimple} is based on 3-address IR, \emph{Simple} is a SSA variant of Jimple and \emph{Grimp} is
an aggregated version of Jimple. By default Soot uses Jimple IR for all program analysis. In this
work we will only use Jimple IR for Doop and ignore the pointer analysis implemented in Soot, as
it natively provides only one fixed configuration, a context-insensitive, 1-object-sensitive analysis, but our evaluation uses 4 different configurations.\\
%\subsection{Benchmark}
\textbf{DaCapo} DaCapo~\cite{dacapo} is a benchmark comprising widely-used open source Java
applications such as Eclipse and Apache FOP.

% \subsubsection{PointBench} \todo{not sure this belongs here}We have developed a micro-benchmark PointBench, consisting of Java
% applications using all possible features of the Java programming language \todo{check}except for reflection and native code. It consists of 9 Java applications.

%% file: sections/methodology.tex
%!TEX root = ../PointEval.tex
%In this work, we study two full pointer analysis frameworks, Doop and Wala, the former of which leverages the IR of either Wala or Soot. In the sequel, we describe the environment and techniques we intend to study.
%\subsection{Environment}
We evaluated Doop and Wala on a server running \emph{Cent\-OS~6.9} on a Xeon processor
with 32 cores and 256GB RAM. Choosing a parallel framework does not benefit
Wala-%and Soot
based pointer analysis because of the sequential nature of their implementation.
However, we restrict Doop's parallelism option to 4 cores, as
this is the configuration that previous Doop publications have been
reporting~\cite{yannis-doop-souffle-2017,YannisOOPSLA2009}. % to have a common basis for comparison of execution time with Wala and Soot.
We use Java8 JRE and assigned 12~GB of heap and 12~GB of stack
space for all experiments. We use the \emph{Doop version 4.12.3} and \emph{Wala version 1.5.0}
%version and \emph{Soot3.3.0} for our evaluation.

%Write about the reflection part
%   1. Reflection analysis used in Doop - TAMIFLEX
%   2. Reflection analysis used in Wala - String resolutio 
\paragraph{\textbf{Reflection}}
The latest version of the DaCapo benchmark uses reflection to invoke other benchmarks, limiting the use of DaCapo
without a sophisticated and precise reflection analysis. We leverage the standard
reflection handling method available in the frameworks. Doop %and Soot
uses Tamiflex~\cite{tamiflex2011Bodden} as the standard reflection analysis. Tamiflex runs as a \emph{java-agent} and logs the types of all objects created through reflective calls. This logged information is
fed into the pointer analysis. %For Soot, we use the standard reflection provided by Tamiflex.
For Wala, we use the the option \texttt{ReflectionOptions.FULL}, which statically resolves the strings passed to reflective calls~\cite{wala-reflection}.

\paragraph{\textbf{Pointer Analysis Techniques Compared}}
For heap abstraction we leverage \emph{allocation site abstraction}, i.e., each object is uniquely denoted by its allocation site, for Doop, and \emph{type-based abstraction} for Wala, as the standard in that framework. In this paper we limit our study to \emph{call-site-sensitive} and \emph{object-sensitive} function invocation abstraction. To be precise, we compare the following techniques of pointer analysis: 
\begin{enumerate}
    \item Call-Site Sensitive : We compare a \emph{one call-site-sensitive} and
    \emph{two call-site-sensitive} pointer analysis. 
    \item Object Sensitive : We compare a \emph{one object-sensitive} and
    \emph{two object-sensitive} pointer analysis. 
\end{enumerate}

We deliberately ignore \emph{type-sensitivity}, as it is a special case of object-sensitivity that only distinguishes
between target objects of the same type. Thus, we try to answer the following research questions
\begin{itemize}\label{methodology:rq}
    \item \textbf{RQ1 \label{question:rq1}} How do our pointer analysis results compare to results that are already published?
    \item \textbf{RQ2 \label{question:rq2}} How do the pointer analysis results change with differing IRs?
    \item \textbf{RQ3 \label{question:rq3}} How do the pointer analysis results compare for our microbenchmark?
    \item \textbf{RQ4 \label{question:rq4}} What are the strengths, weaknesses and usage scenarios of each tool?
\end{itemize}

%% To write
%% We choose to use Geometric PTA which is is supported by Soot. It does a full pointer analysis

%% file: sections/evaluation.tex
%!TEX root = ../PointEval.tex
\subsection{Differences in Class Hierarchy}
\begin{table}[tb]
    \centering
    \begin{adjustbox}{width=0.65\linewidth}
	\small
    \begin{tabular}{|l|r|r|r|r|}
    \hline
    \multirow{2}{*}{Benchmark} & \multicolumn{2}{c|}{\#classes analyzed} & \multirow{2}{*}{\bf Soot exclusive} & \multirow{2}{*}{\bf Wala exclusive} 
    \\ \cline{2-3}
    & Wala & Soot & &   \\ \hline
    Avrora & 21997 & 9204 & 0 & 12793 \\ \hline
    Batik & 23461 & 10739 & 12 & 12734 \\ \hline
    Eclipse & 25718 & 9813 & 62 & 15967 \\ \hline
    H2 & 21007 & 8042 & 1 & 12966 \\ \hline
    Jython & 23323 & 10411 & 2 & 12914 \\ \hline
    Lusearch & 20469 & 4671 & 53 & 15851 \\ \hline
    Luindex & 20479 & 4681 & 53 & 15851 \\ \hline
    PMD & 21315 & 8517 & 1 & 12799 \\ \hline
    SunFlow & 20677 & 7847 & 0 & 12830 \\ \hline
    Tradebeans & 20658 & 3951 & 0 & 16707 \\ \hline
    Xalan & 22688 & 10164 & 0 & 12524 \\ \hline
    
    \end{tabular}
    \end{adjustbox}
    \caption{Table showing the difference in classes considered by Soot and Wala.} %Fourth column shows
   % the extra classes loaded by Soot and fifth column shows the extra classes loaded by Wala.}
    \label{table:differece-classes-wala}
\end{table}

For a whole program analysis, static analysis frameworks consider binary files~(\verb|jar| files or
\verb|class| files), runtime libraries (e.g. \path{rt.jar}, the Java runtime library) and
dependent libraries as input. They then build a class hierarchy based on the classes present in
both the binary files and the libraries. We notice differences in the set of classes contained
in the class hierarchy determined by each framework and decided to investigate these differences
further to understand the possible implications on precision and soundness of these frameworks. It should be noted that we did \emph{not} manually configure Wala to exclude certain
classes, which is a common trick to improve scalability (potentially) at the expense of some soundness. Table~\ref{table:differece-classes-wala} lists the differences between the classes
considered in the Soot and Wala frameworks, which serve as input for Doop and Wala. Consider the case of the class hierarchy defined by Wala for
the benchmark \emph{Avrora}, which loads 1042 classes from the package \path{com.sun.corba} that are not present in the class hierarchy determined by Soot. In this case the classes in the class hierarchy for
Soot are strictly a subset of those in Wala. However, there are cases where a distinct set of classes
is lacking exclusively by either one framework. For example, in the benchmark
\emph{Lusearch}, Soot loads classes from the package \path{javax.crypto} which are not present in the
class hierarchy by \emph{Wala}. In the case of PMD and Sunflow, Soot additionally loads the
class \path{org.junit.Test$None} and \path{org.apache.lucene.store.Lock$With} respectively. We also
notice that 3,890 classes are common to all benchmarks in Soot, and 19,905 classes in Wala. These classes mostly represent the essential parts of the runtime library.
% Static analysis frameworks loads various library
% classes in addition to the ``application'' level classes. We analyze which classes does each 
% framework loads for each benchmark in \emph{DaCapo-bach} and the corresponding relationships. Table~\ref{table:differece-classes-wala}
% lists differnce in the class loaded by both static analysis frameworks i.e. Soot and Wala. In many cases, for e.g. Avrora, Sunflow, classes load by Soot is a subset of classes 
In what follows, we discuss the results of our evaluation on the \emph{DaCapo} benchmarks with
respect to the research questions mentioned in section~\ref{methodology:rq}.

\paragraph*{\textbf{RQ1: How do our pointer analysis results compare to results that are already published?}}

%\subsubsection{Doop}
We ran \emph{Doop} on each benchmark application for 7 hours with varying levels of context-sensitivity. Our
benchmarks are based on the \emph{DaCapo-bach} and \emph{DaCapo-2006} versions of the \emph{DaCapo}
benchmark. \emph{DaCapo-bach} is the latest version of the \emph{DaCapo} benchmark. In subsequent
paragraphs, we discuss the runtime performance and points-to set statistics of our evaluation.

\begin{table}
\centering
 \begin{adjustbox}{width=0.7\linewidth}
	\small
\begin{tabular}{|l|r|r|r|r|r|r|r|r|}
\hline
Analysis & \multicolumn{2}{c|}{1-Call} & \multicolumn{2}{c|}{2-Call} & \multicolumn{2}{c|}{1-Object} & \multicolumn{2}{c|}{2-Object} \\ \hline
& Wala & Jimple & Wala & Jimple & Wala & Jimple & Wala & Jimple \\ \hline
Avrora & 337 & 202 & 17,504 & 15,522 & 238 & 98 & 237 & 54 \\ \hline
Batik & 410 & 246 & 14,853 & 13,045 & 239 & 112 & 274 & 112 \\ \hline
Eclipse & 1208 & 1043 & Timeout & Timeout & 951 & 803 & Timeout & Timeout \\ \hline
H2 & 352 & 188 & 13,379 & 12,994 & 240 & 103 & 234 & 98 \\ \hline
Jython & 460 & 626 & 626 & 22,988 & 573 & 458 & Timeout & Timeout \\ \hline
Luindex & 330 & 132 & 14,182 & 12,770 & 224 & 58 & 234 & 59 \\ \hline
Lusearch & 328 & 115 & 13,841 & 12,987 & 222 & 113 & 242 & 57 \\ \hline
PMD & 359 & 278 & 22,520 & 21,480 & 265 & 117 & 257 & 117 \\ \hline
Sunflow & 357 & 234 & Timeout & 20,204 & 229 & 115 & 256 & 123 \\ \hline
Tradebeans & 336 & 202 & 14,373 & 13,132 & 221 & 51 & 240 & 54 \\ \hline
Xalan & 422 & 849 & 19,504 & 15,077 & 296 & 113 & 290 & 112 \\ \hline
\end{tabular}
\end{adjustbox}
\caption{Analysis runtime (in seconds) for Doop on DaCapo-bach benchmarks}
\label{table:runtime-doop-dacapo-bach}
\end{table}
\begin{table}
    \centering
    \begin{adjustbox}{width=0.5\linewidth}
	\small
\begin{tabular}{|l|r|r|r|r|}
\hline
Benchmark & 1-Call & 2-Call & 1-Object & 2-Object \\ \hline
Antlr & 107 & 380 & 82 & 79 \\ \hline
Bloat & 95 & 3678 & 101 & 2345 \\ \hline
Chart & 143 & 17956 & 119 & 123 \\ \hline
Eclipse & 101 & 17373 & 96 &  \\ \hline
Fop & 98 & 10680 & 100 & 81 \\ \hline
HSQLDB & 80 & 96 & 84 & 75 \\ \hline
Jython & 596 & Timeout & 7644 &  \\ \hline
Luindex & 62 & 12867 & 54 & 57 \\ \hline
Lusearch & 46 & 49 & 45 & 57 \\ \hline
PMD & 136 & 12385 & 104 & 120 \\ \hline
Xalan & 70 & 84 & 79 & 77 \\ \hline
\end{tabular}
\end{adjustbox}
\caption{Analysis runtime (in seconds) with Doop on DaCapo-2006. Timeout denotes that the analysis did not terminate within 7 hours.}
\label{table:runtime-doop-dacapo-2006}
\vspace{-3em}
\end{table}
Table ~\ref{table:runtime-doop-dacapo-bach} show the timings of our evaluation on the \emph{DaCapo-bach} benchmarks. We observe that a 1-call-site
sensitive analysis of \emph{DaCapo-bach} with \emph{Doop} terminates within 19 minutes
(cf.~Table.~\ref{table:runtime-doop-dacapo-bach}). Increasing the precision to a
2-call-site analysis, Eclipse and Sunflow do not terminate within 7 hours, while the other benchmarks
terminate within 6.5 hours, even for large benchmarks like Jython. If
we switch the contexts from call-sites to receiver objects (i.e. from call-site
sensitive to object-sensitive), the results differ. The analysis of 9 (out of 11) benchmarks terminates within 5 minutes (!) for a highly precise analysis such
as \emph{2-object-sensitive} (cf.~Table.~\ref{table:runtime-doop-dacapo-bach}).
Interestingly, 1-object sensitive analysis takes longer to terminate, with all benchmarks terminating
within 17 minutes. This evaluation agrees with the popular notion that an object-sensitive analysis
is more scalable than a call-site sensitive analysis. 

For comparison with published results, we compare the runtime statistics for \emph{DaCapo-2006} with
the results from Antoniadis et. al.~\cite{yannis-doop-souffle-2017}. This work also leverages Soufflé, a Datalog engine, in contrast to the earlier work by Branvenboer et.
al.~\cite{YannisOOPSLA2009}, which uses the LogicBlox Datalog engine.
Table~\ref{table:runtime-doop-dacapo-2006} shows the runtimes of all benchmarks taken from DaCapo-2006. In
our setup, we notice differences from the evaluation results in Antoniadis et.
al. For example, the runtime of all benchmarks in our analysis varied
within a range of $\pm20\%$ from their reported numbers for a 1-call-site analysis. For the 2-object-sensitive analysis,
\emph{Jython} and \emph{Eclipse} did not terminate in our experiments, while in Antoniadis et.
al. it did not terminate for Jython only. For other benchmarks, we observe a $\pm12\%$ difference in runtime compared to Antoniadis et. al. We are unable to compare the other analysis results due to lacking numbers in Antoniadis et. al.
\paragraph*{Points-To Statistics - DaCapo 2006}
\begin{table}
\centering
\begin{tabular}{|l|l|l|l|l|l|}
\hline
\multirow{2}{*}{Benchmark} & \multirow{2}{*}{Analysis} & \multicolumn{2}{|c|}{Original} & \multicolumn{2}{|c|}{Our Evaluation}  \\\cline{3-6}
& & HO & Average & HO & Average \\ \hline
Antlr & 1-call & 4.9M & 31 & 2.4M & 15.15 \\ 
& 2-call+1H & 48M & 84 &  16M & 30.5 \\
& 1-obj+H & 25M & 86 & 1M & 11.1 \\
& 2-obj+1H & 7.8M & 8 & 1M & 7.8 \\ \hline 
Chart & 1-call & 18M & 66 & 4M  & 14.8 \\  
&  2-call+1H & 202M & 173 & 16M & 30.5 \\ 
& 1-obj & 81M & 123 & 70M & 30 \\ 
& 2-obj+1H & 24M & 7 & 1.5M & 4.4 \\ \hline
PMD & 1-call & 5.8M & 31 & 1.3M & 8.56 \\ 
& 2-call+1H & 65M & 94 & 55M & 29.3 \\ 
& 1-obj &  5.2M & 15  & 42K & 5.03 \\
& 2-obj+1H & 7.4M & 7 & 1.99M & 11.6 \\  \hline
Xalan & 1-call & 7.5M & 35 & 47K & 2.1 \\ 
& 2-call+1H & 78M & 88 & 107K & 2.02 \\ 
& 1-obj & 19M  & 30 & 3M & 4.4 \\
& 2-obj+1H & X & X & 40.7K & 1.8 \\  \hline

\end{tabular}
\caption{Comparison of our evaluation of Doop on DaCapo-2006 with the Bravenboaer
et.al.~\cite{YannisOOPSLA2009}. Third and fifth column lists the number of Heap Objects(HO).}
\label{table:dacapo-2006-comparison}
\end{table}

% \begin{table}[tb]
% \centering
% \begin{tabular}{|l|r|r|r|r|}
% \hline
% {\textbf{Benchmark}} & {\centering \textbf{Points-To Set Size}}
% & {\textbf{\centering \#Local Variables}} & \textbf{Average} & {\centering \textbf{Call Graph Edges}} \\ 
% \hline
% \hline
% Avrora & 155,206 & 97,562 & 1.59 & 222,500  \\ \hline
% Batik & 785,552 &   89,405 & 8.79  & 213,518 \\ \hline
% Eclipse & 62,340,549 & 1,249,913  & 49.88  & 168,617  \\ \hline
% H2 & 659,082 & 78,977 & 8.35  & 208,491 \\ \hline
% Jython & 8,085,401 & 293,586 & 27.54  & 390,124 \\ \hline
% Luindex & 681,169  & 85,982 & 7.92   & 210,891\\ \hline
% Lusearch & 576,635 & 61,709 & 9.34 &  196,654  \\ \hline
% PMD & 1,107,518 &   118,710 & 9.30 &  234,706  \\ \hline
% Sunflow & 2,753,127 &  235,249 & 11.7 &  325,326  \\ \hline
% Tradebeans & 563,840 & 60,457 & 9.33 &  196,074  \\ \hline
% Xalan & 1,952,825 &   150,204 & 13.0 &  260,421 \\ \hline
% \end{tabular}
% \caption{Points-to set statistics for Doop used on Wala-IR with 1-Call-site sensitive. Points-To set corresponds to the Var Points-to size.}
% \label{table:doop-wala-one-call-site}
% \end{table} 

\begin{table}[h]
\centering
\begin{tabular}{|l|r|r|r|r|}
\hline
{\textbf{Benchmark}} & {\centering \textbf{Points-To Set Size}}
& {\textbf{\centering \#Local Variables}} & \textbf{Average} & {\centering \textbf{Call Graph Edges}} \\ 
\hline
\hline
Avrora & 933,979 & 126,193 & 7.4  & 220,424\\ \hline
Batik & 973,778 & 120,162 & 8.1  & 213,518\\ \hline
Eclipse & 68,584,518 & 1,549,816 & 44.25  & 2,612,913\\ \hline
H2 & 719,486 & 100,751 & 7.14  & 205,443\\ \hline
Jython & 10,520,213 & 367,517 & 28.63  & 387,718\\ \hline
Luindex & 804,708 & 29,451 & 27.32  & 210,891\\ \hline
Lusearch & 656,531 & 84,798 & 7.74  & 196,654\\ \hline
PMD & 1,443,929 & 157,985 & 9.14  & 234,706\\ \hline
Sunflow & 2,753,127 & 303,181 & 9.08  & 325,325\\ \hline
Tradebeans & 933,979 & 126,193 & 7.4  & 220,424\\ \hline
Xalan & 1,346,974 & 148,926 & 9.04 & 226,249\\ \hline
\end{tabular}
\caption{Points-to set statistics for Doop used on Jimple-IR with 1-Call-site sensitive. Points-To set corresponds to the Var Points-to size.}
\label{table:doop-jimple-one-call-site}
\end{table}

% \begin{table}[tb]
% \centering
% \begin{tabular}{|l|r|r|r|r|}
% \hline
% {\textbf{Benchmark}} & {\centering \textbf{Points-To Set Size}}
% & {\textbf{\centering \#Local Variables}} & \textbf{Average} & {\centering \textbf{Call Graph Edges}} \\ 
% \hline
% \hline
% Avrora & 50,089,915 & 1,299,728 & 38.54 &  9,205,732  \\ \hline
% Batik & 48,177,952 & 1,252,085 &  38.48 &  8,952,657  \\ \hline
% Eclipse & \multicolumn{4}{c|}{Timeout} \\ \hline
% H2 & 47,619,035 & 1,206,436 & 39.47 & 8,918,393  \\ \hline
% Jython & 134,841,875 & 2,270,371 & 59.39  & 10,007,513  \\ \hline
% Luindex & 47,580,093 & 1,220,734 & 38.98  & 8,923,019  \\ \hline
% Lusearch & 47,260,769 & 1,159,635 & 40.75  & 8,888,448  \\ \hline
% PMD & 50,674,534 & 1,336,501 & 37.92  & 9,109,482  \\ \hline
% Sunflow & \multicolumn{4}{c|}{Timeout} \\ \hline
% Tradebeans & 47,258,844 & 1,158,157 & 40.81  & 8,887,369  \\ \hline
% Xalan & 57,190,905 & 1,482,895 & 38.57  & 9,233,434  \\ \hline

% \end{tabular}
% \caption{Results for Doop used on Wala IR with 2-Call-site sensitive+Heap. Points-To set corresponds to the Var Points-to size. Timeout means that the analysis did not terminate within 7 hours.}
% \label{table:doop-wala-two-call-site}
% \end{table}

\begin{table}[tb]
\centering
\begin{tabular}{|l|r|r|r|r|}
\hline
{\textbf{Benchmark}} & {\centering \textbf{Points-To Set Size}}
& {\textbf{\centering \#Local Variables}} & \textbf{Average} & {\centering \textbf{Call Graph Edges}} \\ 
\hline
\hline
Avrora & 53,747,799 & 1,859,506 & 28.90  & 9,205,732\\ \hline
Batik & 51,730,071 & 1,795,267 & 28.81  & 8,952,657\\ \hline
Eclipse & \multicolumn{4}{c|}{Timeout} \\ \hline
H2 & 50,181,943 & 1,721,754 & 29.15  & 8,918,393\\ \hline
Jython & 178,275,063 & 3,018,413 & 59.06  & 10,007,513\\ \hline
Luindex & 50,515,368 & 1,761,389 & 28.68  & 8,923,019\\ \hline
Lusearch & 50,042,431 & 1,681,072 & 29.77  & 8,888,448\\ \hline
PMD & 55,509,783 & 1,909,870 & 29.06  & 9,109,482\\ \hline
Sunflow & 72,312,504 & 2,616,580 & 27.64  & 9,647,292\\ \hline
Tradebeans & 50,041,773 & 1,680,536 & 29.78  & 8,887,369\\ \hline
Xalan & 54,378,426 & 1,862,513 & 29.20 & 9,233,434\\ \hline
\end{tabular}
\caption{Results for Doop used on Jimple IR with 2-Call-site sensitive+Heap. Points-To set corresponds to the Var Points-to size. Timeout means that the analysis did not terminate within 7 hours.}
\label{table:doop-jimple-two-call-site}
\end{table}

\begin{table}[tb]
\centering
\begin{tabular}{|l|r|r|r|r|}
\hline
{\textbf{Benchmark}} & {\centering \textbf{Points-To Set Size}}
& {\textbf{\centering \#Local Variables}} & \textbf{Average} & {\centering \textbf{Call Graph Edges}} \\ 
\hline
\hline
Avrora & 532,536 & 107,693 & 4.94 & 115,282\\ \hline
Batik & 395,140 & 28,505 & 13.86  & 111,476\\ \hline
Eclipse & 60,848,330 & 1,395,999 & 43.59  & 7,778,708\\ \hline
H2 & 285,190 & 1,721,754 & 0.17  & 77,663\\ \hline
Jython & 33,065,947 & 773,400 & 42.75  & 6,172,967\\ \hline
Luindex & 398,422 & 88,100 & 4.52  & 99,232\\ \hline
Lusearch & 223,105 & 61,931 & 3.6  & 79,782\\ \hline
PMD & 914,734 & 39,776 & 23.0 & 158,688\\ \hline
Sunflow & 1,790,548 & 229,033 & 7.82 & 224,478  \\ \hline
Tradebeans & 222,737 & 61,688 & 3.61  & 79,668\\ \hline
Xalan & 1,408,129 & 219,455 & 6.42 & 306,364\\ \hline
\end{tabular}
\caption{Results for Doop used on Jimple IR with 1-object-sensitive. Points-To set corresponds to the Var Points-to size. Timeout means that the analysis did not terminate within 7 hours.}
\label{table:doop-jimple-one-object}
\end{table}

\begin{table}[tb]
\centering
\begin{tabular}{|l|r|r|r|r|}
\hline
{\textbf{Benchmark}} & {\centering \textbf{Points-To Set Size}}
& {\textbf{\centering \#Local Variables}} & \textbf{Average} & {\centering \textbf{Call Graph Edges}} \\ 
\hline
\hline
Avrora & 503,918 & 163,757 & 3.08 & 151,317\\ \hline
Batik & 383,681 & 131,100 & 2.93 & 682,017\\ \hline
Eclipse & \multicolumn{4}{c|}{Timeout}\\ \hline
H2 & 340,499 & 110,385 & 3.08 & 128,815\\ \hline
Jython & \multicolumn{4}{c|}{Timeout} \\ \hline
Luindex & 415,187 & 139,054 & 2.99 & 130,527\\ \hline
Lusearch & 279,878 & 80,148 & 3.49 & 103,493\\ \hline
PMD & 858,605 & 220,119 & 3.90 & 302,659\\ \hline
Sunflow & 1,146,726 & 302,610 & 3.79 & 345,422\\ \hline
Tradebeans & 279,830 & 79,941 & 3.50 & 103,402\\ \hline
Xalan & 1,912,526 & 416,651 & 4.59 & 682,017\\ \hline
\end{tabular}

\caption{Results for Doop used on Jimple IR with 2-object-sensitive. Points-To set corresponds to the Var Points-to size. Timeout means that the analysis did not terminate within 7 hours.}
\label{table:doop-jimple-two-object}
\end{table}

To compare the statistics related to points-to set sizes, we base our comparison on the paper on
Doop~\cite{YannisOOPSLA2009} (referred to as DOOP-1), as
the recent paper~\cite{yannis-doop-souffle-2017} gives no information on points-to statistics. The baseline
paper uses \emph{DaCapo-2006} for evaluation. We notice differences in our evaluation from DOOP-1. Consider the case of Antlr, where
DOOP-1 mentions 4.9M objects with an average of 31, whereas our evaluation only reports 2.4M with an
average of 15.15. Similarly, for a 2-call-site sensitive analysis, the average points-to set size of
PMD reduces to 29.06 from 94, and for Xalan it reduces from 65M to 29.20. One major reason for the sharp reduction in points-to set
size might be the use of more sophisticated resolution strategies for reflection, using approaches such as
\emph{TamiFlex}~\cite{tamiflex2011Bodden}. Analysis without \emph{TamiFlex} reports more heap objects than with \emph{TamiFlex}. \emph{Tami\-Flex} logs the reflective calls during a dynamic analysis and represents the logged reflective call as a regular Java call in the call graph. This reduces the
search space for a pointer analysis by filtering some of the infeasible paths, but may also lead to unsound results if a target method is not encountered during the instrumented runtime execution. 

\paragraph*{Points-to statistics - DaCapo bach}

Note that in the previous section we could only compare the evaluation results based on \emph{Jimple IR},
because to the best of our knowledge the evaluation results with \emph{Wala IR} are not available in the
literature. In the sequel, we extend the evaluation to a newer version of the benchmark, i.e.,
\emph{DaCapo-bach}. With these results we are able to compare different analyses based on the underlying IR (see section~\ref{section:rq2}). In
the following, we discuss the comparison of our results in Jimple IR.
\begin{itemize}[nosep]
    \item \textit{1-call-site sensitive vs. 2-call-site sensitive:}
    Table~{\ref{table:doop-jimple-one-call-site}},{\ref{table:doop-jimple-two-call-site}} lists the
    results of our evaluation for call-site sensitivity. Contrary to the popular belief that the
    average size of points-to sets should decrease with increasing levels of context-sensitivity, we
    observe that the average increases. However, this
    aligns with the results by Branvenboer \etal~\cite{YannisOOPSLA2009} where the average number of
    heap objects referred to by a variable increases when increasing the level of sensitivity. This may be due to the fact that the same object is represented by multiple contexts for one allocation site (heap sensitivity).
    \item \textit{1-object sensitive vs. 2-object-sensitive:}
    Table~\ref{table:doop-jimple-one-object},\ref{table:doop-jimple-two-object} list the results of
    our evaluation for object-sensitivity. We observe that the average in this case, decreases when
    increasing the level of sensitivity. For Batik, the average even decreases sharply by a factor
    of 6. However, in some cases, such as Eclipse and Jython, it does not terminate within 7 hours.
    This behavior is expected as a higher level of context-sensitivity yields a more precise
    analysis.
    % \item \textit{Call-site sensitive v/s Object sensitive} To compare different context-sensitivities, we compare call-site sensitivity with object-sensitivity per level of context-sensitivity. %TODO
\end{itemize}
% In case of Wala IR, our results are similar to those discussed in Doop. Table~\ref{table:doop-wala-one-call-site},\ref{table:doop-wala-two-call-site},\ref{table:doop-wala-one-object}, \ref{table:doop-wala-two-object} shows statistical information of the evaluation of Doop using Wala IR.

\subsubsection{Wala}
We analyzed the pointer analysis algorithm available in Wala. For call-site sensitive analysis, we
use the \verb|makeNCFABuilder| method, which invokes a call-site sensitive analysis parameterized
over the depth of the call-string. For an object-sensitive analysis, Wala provides a method allowing
infinite context-sensitivity for objects of the Java Collection classes. In the subsequent paragraphs,
we discuss our observations with respect to points-to set and runtime statistics.

\begin{figure}[]
    \centering
    \includegraphics[scale=0.4]{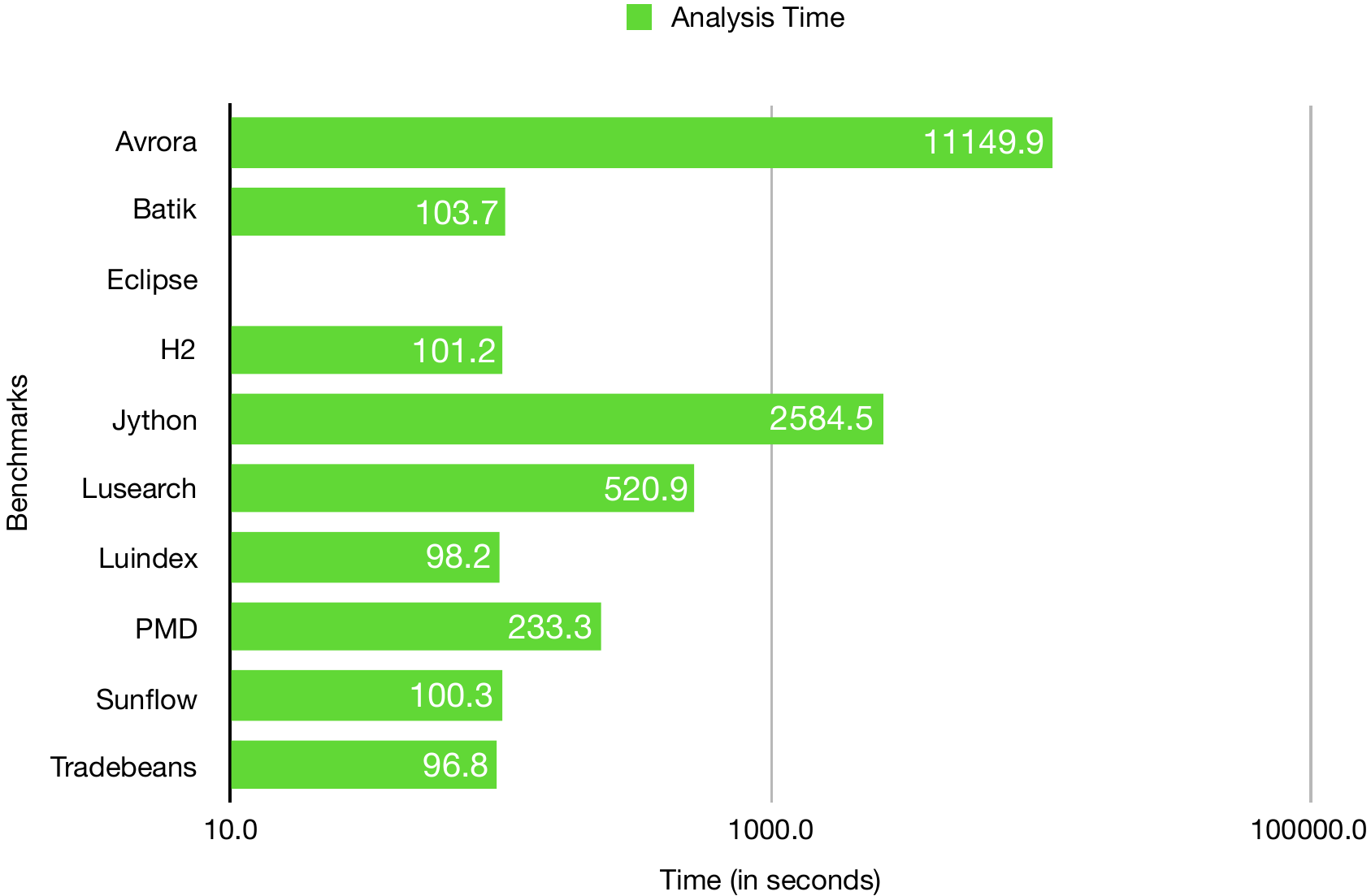}
    \caption{1-call site analysis with Wala Framework}
    \label{figure:wala-one-call-site}
    \vspace{-2em}
\end{figure}
    
\paragraph{Runtime Performance} We observe that the a 1 call-site sensitive analysis in Wala
analysis terminates in at most 3 hours. Interestingly, Wala analysis fails to terminate within 7
hours for Eclipse while \emph{Doop} is fast in analyzing it. However, a 2 call-site sensitive
analysis in Wala fails to terminate within 7 hours for all programs of the DaCapo-bach benchmark
suite. Figure \ref{figure:wala-one-call-site} shows the runtime performance of Wala on the DaCapo
benchmark.

\begin{table}[tb]
    \centering
    \begin{tabular}{|l|r|r|r|r|}
    \hline
    {\textbf{Benchmark}} & {\centering \textbf{Points-To Set Size}}
    & {\textbf{\centering \#Local Variables}} & \textbf{Average} & {\centering \textbf{Call Graph Edges}} \\
    \hline \hline
    Avrora & 207,203,385 & 1,761,087 & 117.66 & 1,064,110  \\ \hline
    Batik & 23,832,702 & 763,298 & 31.22 & 428,517  \\ \hline
    % Eclipse & 402,003,978 & 2,970,559 & 135.33 & 2,629,119  \\ \hline
    Eclipse & \multicolumn{4}{c|}{Timeout}  \\ \hline
    H2 & 23,807,180 & 763,115 & 31.20 & 428,381  \\ \hline
    Jython & 148,180,864 & 1,502,402 & 98.63 & 698,408  \\ \hline
    Lusearch & 31,672,174 & 855,490 & 37.02 & 460,065  \\ \hline
    Luindex & 22,597,093 & 741,638 & 30.47 & 419,944  \\ \hline
    PMD & 53,931,113 & 1,033,157 & 52.20 & 592,199  \\ \hline
    Sunflow & 23,807,180 & 763,115 & 31.20 & 428,381  \\ \hline
    Tradebeans & 23,901,529 & 764,324 & 31.27 & 428,988  \\ \hline
    \end{tabular}
    \caption{Results for Wala  1-call-site sensitive analysis on DaCapo-bach. Points-To set corresponds to the Var Points-to size.}
    \label{table:wala-one-call-site}
\end{table}

\paragraph{Points-To set statistics} To the best of our knowledge, there are no published studies
related to pointer analysis in Wala. Therefore, it is difficult to compare our evaluation with
previous studies. In this paper, we highlight the results of our evaluation of Wala on the DaCapo
benchmark. Table~\ref{table:wala-one-call-site} lists the analysis results for the DaCapo-bach
benchmarks. We observe differences in average sizes of points-to sets for the DaCapo benchmarks, e.g.,
in case of Avrora, where Doop using Wala's front-end reports an average of 1.59, but in the
case of Wala, this increases to 117.66. The difference stems from the choice of heap abstraction in
case of Wala and Doop. Doop uses a heap abstraction which models heap allocation as a pair of
$(\mathit{Heap},\mathit{Context})$, where \( \mathit{Heap} \) is the symbolic name for the allocation site and \( \mathit{Context} \) is
the context where the new object allocation is defined.  In contrast, Wala leverages a simple type-based
heap abstraction where objects of the same type are grouped into a single equivalence class. The heap
abstraction leveraged in Doop renders the analysis more precise by defining a more precise heap abstraction,
which also tracks the context information.

\paragraph{Discussion}
Even though Branvenboer et. al. used the LogicBlox solver for Datalog while the recent version uses
Soufflé~\cite{Souffle2019}, we believe this should not impact the points-to set statistics. Based on
our evaluation on \emph{DaCapo-2006}, we notice differences in the points-to set statistics and
therefore, can safely conclude that the evaluation results of the previous research is not
reproducible. To the best of our knowledge, we did not find any results related to evaluation of
\emph{Doop} on \emph{DaCapo-bach}, therefore we could not compare our evaluation with any previous
results. However, our results can still be reused as a benchmark for further static analysis and
also help researchers in making an informed decision when choosing the appropriate analysis for their
research.

% 
%  Research Question 2
% 
\paragraph*{\textbf{RQ2: How do pointer analysis results change with IR?}} \label{section:rq2}
In this section, we compare the analysis results of Doop using Wala IR and Jimple IR as frontends.
We notice differences in the points-to set statistics when choosing different IRs. For instance in
a 1-call-site sensitive analysis, we notice differences in the average of points-to set size for the benchmarks \emph{Avrora} and \emph{Xalan}. Avrora's average is 7.4 and 1.59 for Jimple IR and
Wala IR, respectively. When analyzing with Wala IR, it yields a near perfect average, while with
Jimple IR it is more coarse. However, the average points-to sizes for all other benchmarks are similar (cf.Table~\ref{table:doop-jimple-one-call-site}).
%where we see a difference of 5.8 in the average points-to size for Jimple IR and Wala IR. Except this case, all other benchmarks have similar average points-to set size.
In terms of runtime performance, the 2-call-site analysis of \emph{Sunflow} with Jimple IR terminates,
while with Wala IR it fails to terminate.
% 
% Research Question 3:
% 
\paragraph*{\textbf{RQ3: What are the main strengths and weaknesses of each tool?}}
%In this section we provide our detailed insights of the results, i.e., where these tools
%lack. Is there a common weakness? 
We compare the tools based on the following metrics:
%\begin{itemize}
 %   \item \emph{Scalability}:  Does the analysis scale for large benchmarks?
  %  \item \emph{Precision}: The lower the value of the average points-to set size, the higher is the precision.
   % \item \emph{Usability}: How can the pointer analysis be integrated into a potential client analysis?
%\end{itemize}

\textbf{Scalability} We find the Doop is more scalable than Wala. It is evident from the the
fact that Doop can  analyze the DaCapo benchmarks for highly precise analyses such as 2-call-site sensitive
analysis. Wala fails to terminate in 7 hours for benchmarks such as \emph{Avrora} while Doop
analyzes that benchmark in just 7.5 minutes. Even for various heavy applications such as Eclipse and Chart, Doop analyzes those within 5 hours. Note that Doop models reflection based on Tamiflex, which is precise but potentially unsound.

\textbf{Precision} Pointer analysis with Doop is more precise than with Wala. Considering the
case of Avrora, Doop exhibits a 1.59 ratio of heap allocation sites to local variables with Wala IR
while a native Wala analysis reports a ratio of 117.6. Comparing Doop analysis with Wala IR and
Jimple IR, Wala has lower averages compared to Doop using Jimple IR.

\textbf{Usability} 
Doop, being based on a declarative specification of pointer analysis outputs a set of files
which define the points-to analysis. As a stand-alone tool for pointer analysis, Doop fares well
in that use case. However, it requires some careful preprocessing steps if one wants to integrate
its results into any client analysis. In contrast, Wala is a static
analysis framework written in Java. Including Wala into various higher level analyses is merely an effort of
invoking some functions.

%
% Research Question 4
% 
% \subsection{Case of Avrora}

%% file: sections/evaluation-pb.tex
%!TEX root = ../PointEval.tex
We also devise a microbenchmark \emph{PointBench} that provides ground truth on the expected points-to sets. \emph{PointBench} consists of 10 corner case
programs, which are small but, in a general sense, challenging for static analysis. These
applications are based on simple features of programming languages and exclude \emph{Reflection,
Dynamic Proxies}, and similar dynamic features.

%\subsection{Characteristics of Applications in PointBench}

First, we give a detailed summary about the applications chosen in our microbenchmark,
\emph{PointBench}. We choose various corner cases for pointer analysis. The applications that we
contrive for our study exhibit these characteristics: (1) The application has no new allocations
within a loop, (2) nested calls are deliberately restricted to 2 and (3) new object allocations are
only done in constructors or the \emph{main} function. 

\emph{Array} is an application which creates and allocates a pre-defined number of objects in an
array. \emph{Context} creates nested contexts up to length 2, the maximum length of contexts to
which analyses scale within a few hours. \emph{Interface, Inheritance} use polymorphic or
dynamic binding, solved challenges of pointer analysis. \emph{MainString} uses a library
class \emph{String} and performs some basic string manipulation operations. \emph{MyVector} defines
a vector-like library class. It includes features that are considered difficult to model, such as
an array of objects and nested function calls. \emph{This} leverages getter and setter
methods to access object fields. The source code and related binary files of our benchmark applications are available at \url{http://bit.ly/pointbench}.
%  We use 
% the the java agent library tool \emph{hprof}~\cite{hprof} and run the tool.
% Design of our benchmark
% ensures that the application can be run without any user input. Therefore, we achieve \(100\%\) code
% coverage during runtime and therefore guarantee that the heap objects values obtain with a dynamic
% analysis are correct.

\begin{table}[tb]
    \centering
    \begin{tabular}{|l|l|l|l|l|}
        \hline
            & & \multicolumn{3}{c|}{Heap Objects} \\ \cline{3-5}
        Benchmark Name & LOC & \#Application & \#Library  & Total  \\ \hline
        Array & 12 &  16 & 1264 & 1280 \\ \hline
        Assign & 9 & 2 & 1293 & 1295 \\ \hline
        Context & 28 & 14 & 1261 & 1275 \\ \hline
        Interface & 18 & 4 & 1507 & 1511 \\ \hline
        MainString & 13 &  96 & 1180 & 1276 \\ \hline
        MyVector & 27 & 2 & 1278 & 1280 \\ \hline
        PiNode & 30 & 1 & 1385 & 1386 \\ \hline
        This & 29 & 2 & 1351 & 1353 \\ \hline
        Inheritance & 42 & 3 & 1424 & 1427 \\ \hline

    \end{tabular}
    \caption{Application in microbenchmark \emph{PointBench}. Third columns shows the new allocation sites each program.}
    \label{table:pointbench-summary}
\end{table}

\begin{table*}[tb]
    \adjustbox{max width=\linewidth}{
    \begin{tabular}{|l|l|r|r|r|r|r|r|r|r|r|r|r|r|}
        \hline
        \multirow{2}{*}{Benchmark} & IR & \multicolumn{3}{c|}{1-call-site} & \multicolumn{3}{c|}{2-call-site} & \multicolumn{3}{c|}{1-object} & \multicolumn{3}{c|}{2-object} \\ \cline{3-14}
        & &  \#Pointers & \#Variables & Average & \#Pointers & \#Variables & Average & \#Pointers & \#Variables & Average & \#Pointers & \#Variables & Average \\ \hline
        Array & Soot & 10,406 & 6,036 & 1.72 & 15,765 & 9,775 & 1.61 & 8,704 & 5,206 & 1.67 & 8,657 & 5,207 & 1.66\\
        Array & Wala & 6,886 & 4,249 & 1.62 & 10662 & 6962 & 1.53 & 5,766 & 3,661 & 1.57 & 5,725 & 3,662 & 1.56\\ \hline
        Assign & Soot & 10,602 & 6,109 & 1.74 & 16,132 & 9,937 & 1.62 & 8,834 & 5,269 & 1.68 & 8,787 & 5,270 & 1.67\\
        Assign & Wala & 7,006 & 4,298 & 1.63 & 10889 & 7072 & 1.54 & 5,847 & 3,703 & 1.58 & 5,806 & 3,704 & 1.57\\ \hline
        Context & Soot & 10,468 & 6,078 & 1.72 & 15,841 & 9,834 & 1.61 & 8,770 & 5,250 & 1.67 & 8,723 & 5,251 & 1.66\\
        Context & Wala & 6,922 & 4,271 & 1.62 & 10,468 & 6,994 & 1.50 & 5,802 & 3,683 & 1.58 & 5,761 & 3,684 & 1.56\\ \hline
        Interface & Soot & 10,394 & 6,024 & 1.73 & 15,753 & 9,763 & 1.61 & 8,692 & 5,194 & 1.67 & 8,645 & 5,195 & 1.66\\
        Interface & Wala & 6,877 & 4,240 & 1.62 & 10,653 & 6,953 & 1.53 & 5,757 & 3,652 & 1.58 & 5,716 & 3,653 & 1.56\\ \hline
        String & Soot & 10,407 & 6,041 & 1.72 & 15,766 & 9,780 & 1.61 & 8,700 & 5,202 & 1.67 & 8,653 & 5,203 & 1.66\\
        String & Wala & 6,884 & 4,251 & 1.62 & 10,660 & 6,964 & 1.53 & 5,761 & 3,656 & 1.58 & 5,720 & 3,657 & 1.56\\ \hline
        Vector & Soot & 10,421 & 6,051 & 1.72 & 15,780 & 9,790 & 1.61 & 8,714 & 5,216 & 1.67 & 8,667 & 5,217 & 1.66\\
        Vector & Wala & 6,895 & 4,258 & 1.62 & 10,671 & 6,971 & 1.53 & 5,772 & 3,667 & 1.57 & 5,731 & 3,668 & 1.56\\ \hline
        PiNode & Soot & 10,634 & 6,133 & 1.73 & 16,135 & 9,953 & 1.62 & 8,863 & 5,302 & 1.67 & 8,816 & 5,303 & 1.66\\
        PiNode & Wala & 7,022 & 4,309 & 1.63 & 10,886 & 7,078 & 1.54 & 5,862 & 3,720 & 1.58 & 5,821 & 3,721 & 1.56\\ \hline
        This & Soot & 103,113,005 & 1,246,209 & 82.74 & TO & TO & TO & 50,679,707 & 1,436,719 & 35.27 & 91,019,445 & 2,736,000 & 33.27\\
        This & Wala & TO & TO & TO & TO & TO & TO & TO & TO & TO & TO & TO & TO\\ \hline
        Inheritance & Soot & 106,808,613 & 1,243,730 & 85.88 & TO & TO & TO & 48,781,288 & 1,439,998 & 33.88 & 104,993,647 & 2,754,849 & 38.11\\
        Inheritance & Wala & TO & TO & TO & TO & TO & TO & TO & TO & TO & TO & TO & TO\\ \hline
\end{tabular}}
    \caption{Points-to set statistics for analysis of \emph{PointBench} with Doop. Timeout (TO) denotes that the analysis failed to terminate within 90 minutes.}
    \label{table:precision-pointbench}
\end{table*}
\begin{table*}[tb]
    \adjustbox{max width=\linewidth}{
    \begin{tabular}{|l|l|r|r|r|r|r|r|r|r|r|}
        \hline
         & & & \multicolumn{2}{c|}{1-Call-site} & \multicolumn{2}{c|}{2-Call-Site} & \multicolumn{2}{c|}{1-object-sensitive} & \multicolumn{2}{c|}{2-object-sensitive} \\
         \cline{4-11}
        Benchmark & IR & Heap Objects & Heap Objects & Precision (in \%) & Heap Objects & Precision (in \%) & Heap Objects & Precision (in \%) & Heap Objects & Precision (in \%)\\ \hline
Array & Soot & 1280 & 10,406 & 12.30 & 15,765 & 8.12 & 8,704 & 14.71 & 8,657 & 14.79\\ 
 & Wala & 1280 & 6,886 & 18.59 & 10662 & 12.01 & 5,766 & 22.20 & 5,725 & 22.36\\ \hline
Assign & Soot & 1295 & 10,602 & 12.21 & 16,132 & 8.03 & 8,834 & 14.66 & 8,787 & 14.74\\ 
 & Wala & 1295 & 7,006 & 18.48 & 10889 & 11.89 & 5,847 & 22.15 & 5,806 & 22.30\\ \hline
Context & Soot & 1275 & 10,468 & 12.18 & 15,841 & 8.05 & 8,770 & 14.54 & 8,723 & 14.62\\ 
 & Wala & 1275 & 6,922 & 18.42 & 10,468 & 12.18 & 5,802 & 21.98 & 5,761 & 22.13\\ \hline
Interface & Soot & 1511 & 10,394 & 14.54 & 15,753 & 9.59 & 8,692 & 17.38 & 8,645 & 17.48\\ 
 & Wala & 1511 & 6,877 & 21.97 & 10,653 & 14.18 & 5,757 & 26.25 & 5,716 & 26.43\\ \hline
String & Soot & 1276 & 10,407 & 12.26 & 15,766 & 8.09 & 8,700 & 14.67 & 8,653 & 14.75\\ 
 & Wala & 1276 & 6,884 & 18.54 & 10,660 & 11.97 & 5,761 & 22.15 & 5,720 & 22.31\\ \hline
Vector & Soot & 1280 & 10,421 & 12.28 & 15,780 & 8.11 & 8,714 & 14.69 & 8,667 & 14.77\\ 
 & Wala & 1280 & 6,895 & 18.56 & 10,671 & 12.00 & 5,772 & 22.18 & 5,731 & 22.33\\ \hline
PiNode & Soot & 1386 & 10,634 & 13.03 & 16,135 & 8.59 & 8,863 & 15.64 & 8,816 & 15.72\\ 
 & Wala & 1386 & 7,022 & 19.74 & 10,886 & 12.73 & 5,862 & 23.64 & 5,821 & 23.81\\ \hline
This & Soot & 1353 & 103,113,005 & 0.00 & TO & TO & 50,679,707 & 0.00 & 91,019,445 & 0.00\\ 
 & Wala & 1353 & TO & TO & TO & TO & TO & TO & TO & TO\\ \hline
Inheritance & Soot & 1427 & 106,808,613 & 0.00 & TO & TO & 48,781,288 & 0.00 & 104,993,647 & 0.00\\ 
 & Wala & 1427 & TO & TO & TO & TO & TO & TO & TO & TO\\ \hline
    \end{tabular}
    }
    \caption{Precision statistics on microbenchmark for analysis with Doop}
    \label{table:precision-doop}
\end{table*}

\begin{table}[tb]
    \centering
    \adjustbox{max width=\columnwidth}{
\begin{tabular}{|l|l|l|l|l|l|}
    \hline
    & & \multicolumn{2}{c|}{1-call-site} & \multicolumn{2}{c|}{2-call-site} \\ \cline{3-6}
Benchmark & Heap Objects & Heap Objects & Precision & Heap Objects & Precision \\ \hline
Array & 476 & 52 & US & TO & TO\\ \hline
Assign & 476 & 21,909,022 & 0.00 & 103 & US\\ \hline
Context & 476 & 87 & US & TO & TO\\ \hline
Interface & 479 & 21,909,880 & 0.00 & 53 & US\\ \hline
MainString & 476 & 53 & US & 81 & US\\ \hline
Vector & 476 & 81 & US & TO & TO\\ \hline
PiNode & 478 & 21,909,354 & 0.00 & 88 & US\\ \hline
This & 477 & 86 & US & TO & TO\\ \hline
Inheritance & 478 & TO & TO & TO & TO\\ \hline
\end{tabular}}
\caption{Precision statistics for analysis with Wala. TO denotes that the analysis did not terminate within 90 minutes and US denotes that the number of objects computed by the analysis is less than those computed from runtime information.}
\end{table}

\paragraph{\textbf{Precision Computation}}
    To compare all analyses on a common metric, we use the precision score. Precision is defined as
the $P=\mathit{TP}/(\mathit{TP}+\mathit{FP})$ where $\mathit{TP}$ is the number of true positives i.e. the number of heap objects which are created at
runtime, and $\mathit{FP}$ the false positives, the number of heap objects determined by the analysis that are not
actually created at runtime. It is to be noted that the sum of $\mathit{TP}$ and $\mathit{FP}$ is the total number of heap
objects determined by the analysis to be created at runtime. %, assuming that the analysis is sound.
To measure the heap objects
created at runtime, we use a heap profiler (\emph{hprof}~\cite{hprof}). We then parse the heap logs to determine the information about all allocated objects. We
segregate the objects into two categories: (1) Application level objects, i.e., objects created in the
application code and (2) library objects, i.e., those objects used by the \emph{Java} runtime and system libraries. Table~\ref{table:pointbench-summary} lists the information about
library and application heap objects.

\paragraph{\textbf{Evaluation of PointBench on Doop}}
We evaluated Doop on \emph{PointBench} with call-site sensitive and object sensitive analysis
and restrict the context length to 2. We present the results of our evaluation in
Table~\ref{table:precision-pointbench} and corresponding precision statistics in Table~\ref{table:precision-doop}. To our surprise, with Wala IR, a low precision analysis such
as \emph{1-call-site sensitive} of the microbench \emph{This} did not finish within 90 minutes.
We observe that with Wala IR, the analysis of simple programs (with getter and setter methods, or
inheritance) fails to scale for various low precision analyses. We also computed the precision for
all benchmarks, and found that object sensitive analysis is more precise than a call-site-sensitive
analysis. A 1-object-sensitive analysis gives an average precision 16.68 compared to 13.94 for
1-call-site-sensitive analysis. Similarly, precision of object-sensitive analysis increases on
increasing the level of context-sensitivity, while this is not the case for callsite-sensitivity (cf.~Table~\ref{table:precision-doop}). 
% \todo{where?} shows the precision statistics for our benchmark applications.

\paragraph{\textbf{Pointbench evaluation on Wala}}
Table~\ref{table:precision-pointbench-wala} shows the evaluation results using Wala. We evaluated
the \emph{1-call-site} and \emph{2-call-site} sensitive analysis. Similar to the evaluation on Doop,
we also notice inconsistencies in the pointer statistics. For example, a \emph{1-call-site} analysis
on Wala analyzes a few variables and pointers compared to Doop and has high averages. So, we can
conclude that the analysis in Wala is not as precise as Doop. Again, the imprecision stems from the
choice of heap abstraction. Wala follows a type-based heap abstraction while Doop also stores the
allocation site context along with the symbolic heap allocation. However, to our surprise we also
notice poor runtime performance of Wala analysis where up to \(60\%\) of the benchmark applications
time out for a \emph{2-call-site} analysis. We also evaluate the precision for Wala analysis and
found that the precision is close to 0 in 3 out of 9 cases and unsound in remaining 6 for a
1-call-site sensitive analysis. A 2-call-site analysis was worse where 4 out of 9 benchmarks
terminate with unsound analysis and remaining did not terminate within 90 minutes.

\begin{table}[tb]
    \adjustbox{max width=\columnwidth}{
    \begin{tabular}{|l|l|l|l|l|l|l|}
        \hline
        Benchmark & \multicolumn{3}{c|}{1-call-site} & \multicolumn{3}{c|}{2-call-site} \\ \cline{2-7}
         &  \#Pointers & \#Variables & Average & \#Pointers & \#Variables & Average \\ \hline
         Array & 52 & 46 & 1.13 & TO & TO & TO\\ \hline
         Assign & 21909022 & 733232 & 29.88 & 103 & 112 & 0.92\\ \hline
         Context & 87 & 92 & 0.95 & TO & TO & TO\\ \hline
         Interface & 21909880 & 733265 & 29.88 & 53 & 49 & 1.08\\ \hline
         MainString & 53 & 49 & 1.08 & 81 & 60 & 1.35\\ \hline
         MyVector & 81 & 60 & 1.35 & TO & TO & TO\\ \hline
         PiNode & 21909354 & 733259 & 29.88 & 88 & 88 & 1.00\\ \hline
         This & 86 & 82 & 1.05 & TO & TO & TO\\ \hline
         Vector & TO & TO & TO & TO & TO & TO\\ \hline
        %  inherit & TO & TO & TO & TO & TO & TO\\ \hline
         Inheritance & TO & TO & TO & TO & TO & TO\\ \hline
    \end{tabular}}
    \caption{Points-to set statistics for analysis of \emph{PointBench} with Wala. Timeout (TO) denotes that the analysis that failed to terminate within 90 minutes.}
    \label{table:precision-pointbench-wala}
\end{table}

\paragraph{\bf Discussion}
From our evaluation of the various analyses, we observe that \emph{Doop} outperforms \emph{Wala} in
terms of precision and scalability. While Wala gives a close to zero average for the precision over all
benchmarks that terminate, \emph{Doop} performs better with better average scores of precision. However,
comparing Doop with both Jimple and Wala IR, the \emph{Doop} analysis
framework with Wala IR outperforms the analysis with Jimple IR in terms of precision, in contrast
to our evaluation on the DaCapo benchmarks.

We compare the heap objects created by the pointer analyses in Wala and Doop using Wala IR. Both of the
frameworks load classes from the Java system libraries. However, Wala does not
model certain classes such as \path{sun.nio.cs.StreamDecoder} and \path{java.util.ServiceLoader},
while Doop models these system classes in its analysis. Therefore, for applications in our benchmark
such as \emph{Array, Context} or \emph{MainString}, the number of heap objects is in the orders of 10 while
for Doop it is in the orders of 10,000. However, even after such an extensive modeling of heap allocations that happen
in system libraries, Doop is more scalable than Wala, as a 2-call-site sensitive analysis in Doop terminates within 90 minutes for 7 out of 9 benchmark applications, compared to 4 out of 9 in the case of Wala.

%% file: sections/related_work.tex
%!TEX root = ../PointEval.tex
%This paper focuses on two aspects, (1) reproducibility studies and (2) pointer snalysis.
%Reproducibility studies are an integral aspect in research
%with the objective that experiments should be reproducible and repeatable.
%
%\subsection{Reproducibility Studies in Static Analysis}
Qin et. al. \cite{Julia2018Issta} did a survey on three static taint analyzers for Android, namely,
\textsc{AmanDroid, FlowDroid, DroidSafe}. It highlighted major weakness of existing tools and
suggested methods to fix those. This work is orthogonal to our work because we primarily focus on
Java. Pauck et. al.~\cite{Bodden2018Reprod} validated the results of previously mentioned Android taint
analysis tools. Their findings conclude that the majority of the tools fail to run with the recent version of the Android framework. Again this work is orthogonal to our work because we focus on Java. 

%\subsection{Pointer Analysis}
Another major focus of our work is pointer analysis. As mentioned, pointer analysis has gained attention in the last
decades, and we have seen strides in solving several obstacles. We will give an overview of the most recent developments in the sequel: Bravenboer et. al.~\cite{YannisOOPSLA2009} proposed a declarative specification for pointer
analysis. This paper strengthened the transfer of focus of static analysis designers from boilerplate implementations to specifying the analysis on a high level of abstraction. We
compare our results with their evaluation, especially related to information about points-to sets. One interesting future work this paper mentions is comparing their points-to analysis with those available in Wala. Our work
extends their work by evaluating it on the \emph{DaCapo-bach} benchmark suite, and comparing their analysis with Wala. Antonidis et. al.~\cite{yannis-doop-souffle-2017} ported the Datalog engine used by Doop from
LogicBlox to Soufflé. It resulted in accelerating Doop's runtime performance but their paper lacks an
evaluation for a 2-call-site and 1-object-sensitive analysis. We compare to their reported runtime performance numbers in
our evaluation on \emph{DaCapo-2006}, and we also also extend the evaluation of their paper by
providing an extensive evaluation of various analyses. Späth et. al.~\cite{boomerang} proposed a first microbenchmark for pointer analysis,
\emph{PointerBench}. Their benchmark is used for experiments with their demand-driven pointer
analysis, \textsc{Boomerang}~\cite{boomerang}. However, it lacks comparison with other whole-program pointer
analyses available in literature. Rief et. al.~\cite{Reif2018,Rief2019} proposed a tool chain for analyzing the unsoundness of various
call-graph generation algorithms available in Soot, Wala and Doop. It showed the sources of
unsoundness in these algorithms. This work is complementary to our work because of the
interdependency of pointer analysis and call-graph construction.
% \todo{Write about Geom PA}
%TODO: Add Java-8, Java-9 features 

%% file: sections/conclusion.tex
%!TEX root = ../PointEval.tex
This paper reports on the inconsistencies in the static analysis frameworks Doop and Wala, and shows the differences in precision and runtime performance. The differences in the choice of abstractions and the underlying class hierarchy and call graph even for the same frontend (Wala IR) are subtle and render a detailed comparison challenging. However, in general Doop is faster and more precise than Wala in our experiments, but also harder to integrate into client analyses.

%% file: PointEval.bbl
\begin{thebibliography}{10}

\bibitem{yannis-doop-souffle-2017}
Tony Antoniadis, Konstantinos Triantafyllou, and Yannis Smaragdakis.
\newblock Porting doop to souffl\'e;: A tale of inter-engine portability for
  datalog-based analyses.
\newblock In {\em Proceedings of the 6th ACM SIGPLAN International Workshop on
  State Of the Art in Program Analysis}, SOAP 2017, pages 25--30, New York, NY,
  USA, 2017. ACM.
\newblock URL: \url{http://doi.acm.org/10.1145/3088515.3088522}, \href
  {http://dx.doi.org/10.1145/3088515.3088522}
  {\path{doi:10.1145/3088515.3088522}}.

\bibitem{modular-typestate-Aldrich}
Kevin Bierhoff and Jonathan Aldrich.
\newblock Modular typestate checking of aliased objects.
\newblock In {\em Proceedings of the 22Nd Annual ACM SIGPLAN Conference on
  Object-oriented Programming Systems and Applications}, OOPSLA '07, pages
  301--320, New York, NY, USA, 2007. ACM.
\newblock URL: \url{http://doi.acm.org/10.1145/1297027.1297050}, \href
  {http://dx.doi.org/10.1145/1297027.1297050}
  {\path{doi:10.1145/1297027.1297050}}.

\bibitem{dacapo}
Stephen~M. Blackburn, Robin Garner, Chris Hoffmann, Asjad~M. Khang, Kathryn~S.
  McKinley, Rotem Bentzur, Amer Diwan, Daniel Feinberg, Daniel Frampton,
  Samuel~Z. Guyer, Martin Hirzel, Antony Hosking, Maria Jump, Han Lee,
  J.~Eliot~B. Moss, Aashish Phansalkar, Darko Stefanovi\'{c}, Thomas VanDrunen,
  Daniel von Dincklage, and Ben Wiedermann.
\newblock The dacapo benchmarks: Java benchmarking development and analysis.
\newblock In {\em Proceedings of the 21st Annual ACM SIGPLAN Conference on
  Object-oriented Programming Systems, Languages, and Applications}, OOPSLA
  '06, pages 169--190, New York, NY, USA, 2006. ACM.
\newblock URL: \url{http://doi.acm.org/10.1145/1167473.1167488}, \href
  {http://dx.doi.org/10.1145/1167473.1167488}
  {\path{doi:10.1145/1167473.1167488}}.

\bibitem{tamiflex2011Bodden}
Eric Bodden, Andreas Sewe, Jan Sinschek, Hela Oueslati, and Mira Mezini.
\newblock Taming reflection: Aiding static analysis in the presence of
  reflection and custom class loaders.
\newblock In {\em Proceedings of the 33rd International Conference on Software
  Engineering}, pages 241--250, New York, NY, USA, 2011.

\bibitem{YannisOOPSLA2009}
Martin Bravenboer and Yannis Smaragdakis.
\newblock Strictly declarative specification of sophisticated points-to
  analyses.
\newblock In {\em Proceedings of the 24th ACM SIGPLAN Conference on Object
  Oriented Programming Systems Languages and Applications}, OOPSLA '09, pages
  243--262, New York, NY, USA, 2009. ACM.
\newblock URL: \url{http://doi.acm.org/10.1145/1640089.1640108}, \href
  {http://dx.doi.org/10.1145/1640089.1640108}
  {\path{doi:10.1145/1640089.1640108}}.

\bibitem{hprof}
Oracle Corportation.
\newblock Hprof: A heap/cpu profiling tool, Jan 2019.
\newblock URL:
  \url{https://docs.oracle.com/javase/7/docs/technotes/samples/hprof.html}.

\bibitem{Souffle2019}
Oracle Corportation.
\newblock Souffl\'e, Jan 2019.
\newblock URL: \url{https://souffle-lang.github.io/}.

\bibitem{doop2019}
Doop.
\newblock Doop - framework for java pointer and taint analysis (using p/taint),
  June 2019.
\newblock URL: \url{https://bitbucket.org/yanniss/doop}.

\bibitem{Yannis2017-OOPSLA-PTaint}
Neville Grech and Yannis Smaragdakis.
\newblock P/taint: Unified points-to and taint analysis.
\newblock {\em Proc. ACM Program. Lang.}, 1(OOPSLA):102:1--102:28, October
  2017.
\newblock URL: \url{http://doi.acm.org/10.1145/3133926}, \href
  {http://dx.doi.org/10.1145/3133926} {\path{doi:10.1145/3133926}}.

\bibitem{hind-2001-havent-solved-yet}
Michael Hind.
\newblock Pointer analysis: Haven't we solved this problem yet?
\newblock In {\em Proceedings of the 2001 ACM SIGPLAN-SIGSOFT Workshop on
  Program Analysis for Software Tools and Engineering}, PASTE '01, pages
  54--61, New York, NY, USA, 2001. ACM.
\newblock URL: \url{http://doi.acm.org/10.1145/379605.379665}, \href
  {http://dx.doi.org/10.1145/379605.379665} {\path{doi:10.1145/379605.379665}}.

\bibitem{kanvar2016heap-abstraction}
Vini Kanvar and Uday~P. Khedker.
\newblock Heap abstractions for static analysis.
\newblock {\em ACM Comput. Surv.}, 49(2):29:1--29:47, June 2016.
\newblock URL: \url{http://doi.acm.org/10.1145/2931098}, \href
  {http://dx.doi.org/10.1145/2931098} {\path{doi:10.1145/2931098}}.

\bibitem{landi-undecidable}
William Landi.
\newblock Undecidability of static analysis.
\newblock {\em ACM Lett. Program. Lang. Syst.}, 1(4):323--337, December 1992.
\newblock URL: \url{http://doi.acm.org/10.1145/161494.161501}, \href
  {http://dx.doi.org/10.1145/161494.161501} {\path{doi:10.1145/161494.161501}}.

\bibitem{hybdridroid}
S.~{Lee}, J.~{Dolby}, and S.~{Ryu}.
\newblock Hybridroid: Static analysis framework for android hybrid
  applications.
\newblock In {\em 2016 31st IEEE/ACM International Conference on Automated
  Software Engineering (ASE)}, pages 250--261, Sep. 2016.

\bibitem{Bodden2018Reprod}
Felix Pauck, Eric Bodden, and Heike Wehrheim.
\newblock Do android taint analysis tools keep their promises?
\newblock {\em CoRR}, abs/1804.02903, 2018.
\newblock URL: \url{http://arxiv.org/abs/1804.02903}, \href
  {http://arxiv.org/abs/1804.02903} {\path{arXiv:1804.02903}}.

\bibitem{Julia2018Issta}
Lina Qiu, Yingying Wang, and Julia Rubin.
\newblock Analyzing the analyzers: Flowdroid/iccta, amandroid, and droidsafe.
\newblock In {\em Proceedings of the 27th ACM SIGSOFT International Symposium
  on Software Testing and Analysis}, ISSTA 2018, pages 176--186, New York, NY,
  USA, 2018. ACM.
\newblock URL: \url{http://doi.acm.org/10.1145/3213846.3213873}, \href
  {http://dx.doi.org/10.1145/3213846.3213873}
  {\path{doi:10.1145/3213846.3213873}}.

\bibitem{ramalingam-alias}
G.~Ramalingam.
\newblock The undecidability of aliasing.
\newblock {\em ACM Trans. Program. Lang. Syst.}, 16(5):1467--1471, September
  1994.
\newblock URL: \url{http://doi.acm.org/10.1145/186025.186041}, \href
  {http://dx.doi.org/10.1145/186025.186041} {\path{doi:10.1145/186025.186041}}.

\bibitem{eichberg2018soap}
Michael Reif, Florian K\"{u}bler, Michael Eichberg, and Mira Mezini.
\newblock Systematic evaluation of the unsoundness of call graph construction
  algorithms for java.
\newblock In {\em Companion Proceedings for the ISSTA/ECOOP 2018 Workshops},
  ISSTA '18, pages 107--112, New York, NY, USA, 2018. ACM.
\newblock URL: \url{http://doi.acm.org/10.1145/3236454.3236503}, \href
  {http://dx.doi.org/10.1145/3236454.3236503}
  {\path{doi:10.1145/3236454.3236503}}.

\bibitem{Reif2018}
Michael Reif, Florian K\"{u}bler, Michael Eichberg, and Mira Mezini.
\newblock Systematic evaluation of the unsoundness of call graph construction
  algorithms for java.
\newblock In {\em Companion Proceedings for the ISSTA/ECOOP 2018 Workshops},
  ISSTA '18, pages 107--112, New York, NY, USA, 2018. ACM.
\newblock URL: \url{http://doi.acm.org/10.1145/3236454.3236503}, \href
  {http://dx.doi.org/10.1145/3236454.3236503}
  {\path{doi:10.1145/3236454.3236503}}.

\bibitem{Rief2019}
Michael Reif, Florian Kübler, Michael Eichberg, Dominik Helm, and Mira Mezini.
\newblock {Judge: Identifying, Understanding, and Evaluating Sources of
  Unsoundness in Call Graphs}.
\newblock In {\em {Proceedings of the 28th ACM SIGSOFT International Symposium
  on Software Testing and Analysis (to appear)}}, ISSTA 2019, 2019.
\newblock URL: \url{http://dx.doi.org/10.1145/3293882.3330555}, \href
  {http://dx.doi.org/10.1145/3293882.3330555}
  {\path{doi:10.1145/3293882.3330555}}.

\bibitem{Reps-cs-undecidable}
Thomas Reps.
\newblock Undecidability of context-sensitive data-dependence analysis.
\newblock {\em ACM Trans. Program. Lang. Syst.}, 22(1):162--186, January 2000.
\newblock URL: \url{http://doi.acm.org/10.1145/345099.345137}, \href
  {http://dx.doi.org/10.1145/345099.345137} {\path{doi:10.1145/345099.345137}}.

\bibitem{soot2019}
Soot.
\newblock Soot - a framework for analyzing and transforming java and android
  applications, Jan 2019.
\newblock URL: \url{http://sable.github.io/soot/}.

\bibitem{boomerang}
Johannes Sp{\"a}th, Lisa Nguyen~Quang Do, Karim Ali, and Eric Bodden.
\newblock {Boomerang: Demand-Driven Flow- and Context-Sensitive Pointer
  Analysis for Java}.
\newblock In Shriram Krishnamurthi and Benjamin~S. Lerner, editors, {\em 30th
  European Conference on Object-Oriented Programming (ECOOP 2016)}, volume~56
  of {\em Leibniz International Proceedings in Informatics (LIPIcs)}, pages
  22:1--22:26, Dagstuhl, Germany, 2016. Schloss Dagstuhl--Leibniz-Zentrum fuer
  Informatik.
\newblock URL: \url{http://drops.dagstuhl.de/opus/volltexte/2016/6116}, \href
  {http://dx.doi.org/10.4230/LIPIcs.ECOOP.2016.22}
  {\path{doi:10.4230/LIPIcs.ECOOP.2016.22}}.

\bibitem{Sridharan-cfl}
Manu Sridharan and Rastislav Bod\'{\i}k.
\newblock Refinement-based context-sensitive points-to analysis for java.
\newblock In {\em Proceedings of the 27th ACM SIGPLAN Conference on Programming
  Language Design and Implementation}, PLDI '06, pages 387--400, New York, NY,
  USA, 2006. ACM.
\newblock URL: \url{http://doi.acm.org/10.1145/1133981.1134027}, \href
  {http://dx.doi.org/10.1145/1133981.1134027}
  {\path{doi:10.1145/1133981.1134027}}.

\bibitem{wala-reflection}
Wala.
\newblock Wala analysis options, Jun 2019.
\newblock URL:
  \url{https://github.com/wala/WALA/blob/master/com.ibm.wala.core/src/com/ibm/wala/ipa/callgraph/AnalysisOptions.java}.

\bibitem{wala2019}
WALA.
\newblock Watson libraries for program analysis, Jan 2019.
\newblock URL: \url{http://wala.sourceforge.net/wiki/index.php/Main\_Page}.

\bibitem{datalogWiki2019}
Wikipedia.
\newblock Datalog, Jan 2019.
\newblock URL: \url{https://en.wikipedia.org/wiki/Datalog}.

\end{thebibliography}
